\documentclass[natbib]{article}
\usepackage[]{inputenc} 
\usepackage[T1]{fontenc} 
\usepackage{hyperref}       
\usepackage{url}            
\usepackage{graphicx}%
\usepackage{multirow}%
\usepackage{amsmath,amssymb,amsfonts}%
\usepackage{amsthm}%
\usepackage{mathrsfs}%
\usepackage[title]{appendix}%
\usepackage{xcolor}%
\usepackage{textcomp}%
\usepackage{manyfoot}%
\usepackage{booktabs}%
\usepackage{algorithm}%
\usepackage{algorithmicx}%
\usepackage{algpseudocode}%
\usepackage{listings}%
\usepackage{natbib}

\theoremstyle{thmstyleone}%
\theoremstyle{thmstyletwo}%

\theoremstyle{thmstylethree}%

\newcommand{\normal}{\ensuremath{\mathcal{N}}} 
\newcommand{\setN}{\mathbb{N}} 
\newcommand{\setR}{\mathbb{R}} 
\newcommand{\E}{\ensuremath{\mathbb{E}}} 
\DeclareMathOperator{\Var}{Var} 
\DeclareMathOperator{\tr}{tr} 

\algrenewcommand\algorithmicrequire{\textbf{Input:}} 
\algrenewcommand\algorithmicensure{\textbf{Output:}} 

\begin{document}

\title{Model-Based Clustering of Functional Data \\Via Random Projection Ensembles}

\author{Matteo Mori\footnote{via delle Belle Arti 41, 40126 Bologna (Italy) \newline mailto: \url{matteo.mori8@unibo.it}} \and Laura Anderlucci}

\date{}
\maketitle

\vspace{-1cm}
\begin{center} 
Department of Statistical Sciences -  University of Bologna
\end{center}


\begin{abstract}
Clustering functional data is a challenging task due to intrinsic infinite-dimensionality and the need for stable, data-adaptive partitioning. In this work, we propose a clustering framework based on Random Projections, which simultaneously performs dimensionality reduction and generates multiple stochastic representations of the original functions. Each projection is clustered independently, and the resulting partitions are then aggregated through an ensemble consensus procedure, enhancing robustness and mitigating the influence of any single projection. To focus on the most informative representations, projections are ranked according to clustering quality criteria, and only a selected subset is retained. In particular, we adopt Gaussian Mixture Models as base clusterers and employ the Kullback–Leibler divergence to order the random projections; these choices enable fast computation and eliminate the need to specify the number of clusters a priori. The performance of the proposed methodology is assessed through an extensive simulation study and two real-data applications, one from spectroscopy data for food authentication and one from log-periodograms of speech recording; the obtained results suggest that the proposal represents an effective tool for the clustering of functional data.

\vspace{0.5cm}
\noindent%
\vfill
\end{abstract}

\section{Introduction}\label{into}
In many scientific fields, data are collected in the form of curves, trajectories, or other functional observations that vary continuously over a domain. Functional Data Analysis (FDA) provides a theoretical and methodological framework for treating these inherently infinite-dimensional objects as single entities, enabling the extraction of meaningful structure from complex datasets. Most of the theoretical background on functional data is covered in the seminal book by \cite{ramsay2005functional}; a nice review on the applications of FDA can be found in \cite{ullah2013applications}.
In practice, units are observed and measured over a (possibly large) set of $n\in \setN$ domain points $y_j$, $1\leq j\leq n$ from which the (approximated) function is reconstructed. Specifically, let us consider the following model:
\begin{align}\label{fundata_model}
	\mathbf{y}=x(\mathbf{t})+\mathbf{e},
\end{align}
where $\mathbf{y}$ is the vector of observed values, $\mathbf{t}$ is the vector of equi-spaced (time) points, $x$ is the underlying function and $\mathbf{e}$ is the vector of random errors, whose elements are usually modeled as realization of some exogenous random variable with zero mean. Here $\mathbf{y}$, $x(\mathbf{t})$, $\mathbf{t}$ and $\mathbf{e}$ are all column vectors of length $n$, while in the following $x$ or $x(t)$ denote the underlying function.

In this work, the problem of partitioning the observed functional data $\mathbf{y}$ into a set of homogeneous groups, whose elements exhibit similar behaviour, is addressed. Given that the capability to store and process large amounts of functional data has only recently become feasible, the topic of functional data clustering is a relatively new area of research \citep[for a recent review, see][]{reviewZhang}. Nevertheless, applications have emerged in several scientific fields, from medicine \citep[e.g.,][]{scheipl,baragilly2022clustering,tzeng2018dissimilarity} to environmental sciences \citep[e.g.,][]{hael2024dynamic,villani2024climate}, from economics and finance \citep[e.g.,][]{seo2025clustering} to sport \citep[e.g.,][]{fortuna2018clustering,bouvet2024investigating}, to name a few.

%

Following the taxonomy proposed by \cite{survey} in their review, clustering methods for functional data can be cast into four broad categories: \emph{raw data} methods, \emph{filtering} methods, \emph{adaptive} methods, and \emph{distance-based} methods.

Raw data methods, while typically fast, do not account for the functional nature of the data and can be viewed as extensions of multivariate techniques applied directly to the observed time points or their transformations.

Distance-based methods rely on clustering algorithms defined through specific distances for functional data and, depending on how these distances are computed, may fall into either the raw data or filtering category.

Adaptive methods perform dimension reduction and clustering simultaneously. Within this latter category, several approaches have been proposed. For example, in \cite{bouveyron2011model} the clustering is carried out using a functional latent mixture model, where group-specific low-dimensional subspaces that best represent each cluster are estimated and each group is characterized by its own functional principal component structure. Later, \cite{gattone} proposed a functional version of $K$-means algorithm over a reduced subspace, able to combine dimension reduction and adaptive smoothing within a unified framework. The dimensionality reduction is obtained by constraining the cluster centroids to lie onto a subspace which preserves the maximum amount of discriminative information contained in the original data, while smoothness is enforced through a penalized least squares approach with a smoothing parameter automatically tuned via Generalised Cross-Validation (GCV). Differently, \cite{jacques2013funclust} based their work on an approximation for the density of functional variables that allows to fit a Gaussian mixture model (GMM) over the functional principal component scores. \cite{centofanti2024sparse} proposed a model-based clustering method that works for sparse functional data and that relies on a general functional GMM whose parameters are estimated by maximizing a penalised log-likelihood function. The penalty combines a functional adaptive pairwise fusion term and a roughness penalty to enforce smoothness in clustering.
Recently, \cite{yu2025distance} proposed to improve clustering performance by incorporating derivative information into the distance metric. They define a weighted $L^2$-type distance that combines information from both the original curve and their first derivatives, representing them through functional principal component analysis.

Finally, filtering methods generally follow a two-step procedure: first curves are approximated using basis functions, and then clustering is carried out on the resulting basis coefficients. Within this latter framework, \cite{Chen2015EMClusterpackage} proposed to estimate GMMs on the functional principal component scores to cluster functional data. Later, \cite{martino2019k} presented a functional $K$-means algorithm, where the distance between curves is defined through a metric that generalizes Mahalanobis distance in Hilbert spaces. This distance accounts for both the correlation structure and variability of all functional components, extending the classical Mahalanobis distance to infinite-dimensional settings. Recently, \cite{ren2023multivariate} propose a fast, non-iterative method for clustering both univariate and multivariate functional data using an adaptive density peak detection technique. The approach identifies cluster centers based on local functional density estimates and distances to higher-density neighbours, employing functional $k$-nearest neighbour density estimators and functional principal component representations to handle complex data structures efficiently.

The method proposed here falls in this category, too. Using basis function decomposition, rather than observed raw points, allows to regularize the curves, so to account for measurement errors, and to represent functions in a finite space. However, in order to have an accurate representation of the data, basis function expansions can still lead to a high dimensional parameter space, which prevents from using traditional clustering methods as they can be unstable, computationally unfeasible or simply not applicable \citep{hennig2015handbook}. Main approaches to tackle the high-dimensionality problem in clustering consider either feature selection, where it is assumed that only a subset of the variables is relevant in uncovering the clusters structure, or feature extraction, where the clustering space is found by considering new artificial features, e.g. linear combinations of the original ones.
Random Projections (RPs) are a feature extraction method that have been largely employed for different multivariate analysis tasks, e.g. as a data compression technique \citep{freksen2021introduction}, in supervised \citep{cannings2017random} and unsupervised \citep{fern2003random,anderlucci2022high} classification. For a nice review, see \cite{cannings2021random}. The rational is to project high-dimensional data onto a lower-dimensional subspace using a random matrix, while approximately preserving pairwise distances between observations. This approach is theoretically supported by the Johnson-Lindenstrauss (JL) Lemma (1984), which states that any $N$-point set in a $K$ dimensional space can be linearly projected onto a $d=\mathcal{O}(\log(N)/\varepsilon^2)\ll K$ dimensional space by means of a random matrix $\mathbf{A}$ with orthonormal columns, while preserving pairwise distances with a factor $1\pm\varepsilon, \ \varepsilon\in (0,1)$. More formally, given $\mathcal{D}=\{\mathbf{x}_1, \mathbf{x}_2, \ldots, \mathbf{x}_N\}$, $\mathbf{x}_i\in\setR^K$, $i=1,\ldots,N$ we have that, with high probability over the randomness of $\mathbf{A}$:
\begin{align}\label{JLL}
	(1-\varepsilon)\left \lVert \mathbf{x}_i - \mathbf{x}_j \right \rVert_2\leq\left \lVert \mathbf{A}^T\mathbf{x}_i - \mathbf{A}^T\mathbf{x}_j \right \rVert_2\leq(1+\varepsilon)\left \lVert \mathbf{x}_i - \mathbf{x}_j \right \rVert_2,
\end{align}
where $i,j=1,\ldots,N$ and $\left \lVert \cdot \right \rVert_2$ is the $\mathit{L}_2$ norm \citep{johnson1984extensions}.

Later, \cite{dasgupta2003elementary} provided a new proof for the lemma that relaxes the assumption on the orthonormality of matrix $\mathbf{A}$, extending the result to the case of Gaussian projection matrices, too. By projecting high-dimensional data onto a lower-dimensional subspace using a random matrix that satisfies the lemma, the pairwise distances between observations are approximately preserved.

\begin{figure}[htbp]
	\centering
	\includegraphics[width=0.95\textwidth]{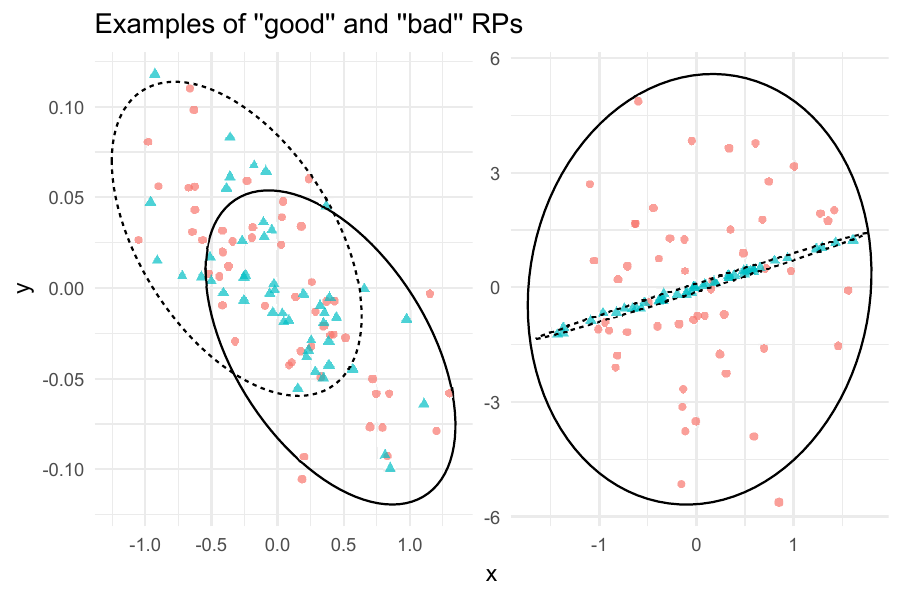}
	\caption{Plots of two RPs. The shape and colour of the points correspond to their true labels, while the ellipses represent the $95\%$ contours levels of the GMM components.}
	\label{good_bad}
\end{figure}

Differently from other dimension reduction methods like principal components, RPs are data oblivious (i.e., they compress data independently from any specific characteristic the data may have). However, an obvious drawback of employing RPs for cluster analysis is their variability: different projections may or may not highlight a grouping structure.
For illustration purposes, consider Figure~\ref{good_bad}, which shows two projections of the same dataset generated in the  simulation study described in detail in Section~\ref{simulations}  (Scenario~\ref{jacques}). In this example, the dimension of the subspace $d$ is set to 2, so as to visualize the projected data in a two-dimensional plot. The ellipses in the figures represent the GMM components fitted to the data, while the points are coloured and shaped according to their cluster memberships. The left panel illustrates the outcome of a random projection that does not reveal a clear group structure, whereas the right panel displays a projection that effectively highlights the underlying cluster separation.

In order to overcome this issue, the results on several different projections can be combined in an ensemble. In general, from $B$ different partitions $\mathbf{G}_b$, a consensus method computes the consensus partition by minimizing a criterion of the form
\begin{align}\label{ensemble_function}
	L(\mathbf{G})=\sum_{b=1}^{B}w_bd(\mathbf{G}_b,\mathbf{G})^p,
\end{align}
for some dissimilarity measure $d(\cdot,\cdot)$, some case weight $w_b$ and some power $p\geq1$ \citep{gordon1999classification}.

In the following, we propose a filtering method that exploits random projections to generate multiple clustering solutions, which are then combined through an ensemble step. The paper is organized as follows: Section~\ref{proposition} describes the proposed model in detail; Section~\ref{simulations} presents the results of simulation studies; Section~\ref{real_data} illustrates two applications to real datasets; and Section~\ref{conclusions} provides a brief discussion and concluding remarks.

\section{The proposed model}\label{proposition}

The goal is to obtain a vector $\mathbf{G}\in\setN^N$ of cluster labels. Denote with matrix $\mathbf{C}\in \setR^{N\times K}$ the collection of the $K$-vectors of basis coefficients for the $N$ curves, obtained either via Generalised Cross-Validation or by other means. Motivated by the JL Lemma, we generate $B$ random matrices $\mathbf{A}_b\in\setR^{K\times d}$ that we use to project the coefficients in $d$-dimensional subspaces, where $d \ll K$. This amounts to compute the matrices
\begin{align}\label{projection_matrixform}
	\mathbf{X}_b=\mathbf{C}\mathbf{A}_b\in\setR^{N\times d},
\end{align}
where a clustering method can be applied to obtain the data partition $\mathbf{G}_b$. Since not all the projections highlight a clear cluster structure, the $B$ vectors of partitions $\mathbf{G}_b$ can be sorted according to a criterion that is able to distinguish the ``useful" partitions from the uninformative ones (this step is discussed in detail in the following). Finally, we obtain the final partition $\mathbf{G}$ by employing an ensemble method on the $B^*$ best partitions. The steps are outlined in Algorithm~\ref{general_algorithm}. This is the general workflow that can be implemented with specific methods at each step. In what follows, we discuss each step of the algorithm in detail.

\begin{algorithm}[H]
	\caption{\textbf{General Outline}}\label{general_algorithm}
	\begin{algorithmic}[1]
		\Require $\mathbf{C} \in \setR^{N\times K}$ matrix of basis coefficients.
		\Ensure $\mathbf{G} \in \setR^N$ vector of group labels.
		\State Set $B$, $B^*$, $d\ll K$;
		\For{b \textbf{from} 1 \textbf{to} B}
		\State \parbox[t]{\dimexpr\textwidth-\leftmargin-\labelsep-\labelwidth}{Generate random matrix $\mathbf{A}_b \in \setR^{K\times d}$\strut}
		\State \parbox[t]{\dimexpr\textwidth-\leftmargin-\labelsep-\labelwidth}{Fit a clustering method on $\mathbf{X}_b=\mathbf{C}\mathbf{A}_b$\strut}
		\State Keep the partition $\mathbf{G}_b$
		\EndFor
		\State Sort the vectors $\mathbf{G}_b$ according to a suitable criterion
		\State Keep the best $B^*$ solutions
		\State Obtain output $\mathbf{G}$ via an ensemble method
		
	\end{algorithmic}
\end{algorithm}

In our implementation, each projected dataset is clustered via a Gaussian Mixture Model. This choice is motivated by the flexibility of the model and, importantly, by the fact that it does not require specifying the number of clusters a priori, while accounting for dependence structures; the Bayesian information criterion is used to select both the optimal number of clusters and the appropriate variance-covariance structure. Note that these may vary across different projected datasets. Using an algorithm that does not require explicit hyperparameter tuning is advantageous, as it avoids the need to determine the best configuration for each of the $B$ model fits or to impose a common specification across all projections. Random matrices can be generated in several ways and different ordering criterion can be used. These choices are discussed in detail in Section~\ref{rp_criterion}, supported by illustrative simulation results.

The last step of the procedure is devoted to the construction of the ensemble, which is a consolidated practice that consists in aggregating multiple clustering results in a single consensus clustering \citep{boongoen2018cluster}. The employed method, that can be found in \cite{dimitriadou2002combination}, is a fixed-point algorithm that obtains soft least squares Euclidean consensus partitions over all soft partitions with a given maximal number of classes. This amounts to minimise the right-hand side of Equation~\ref{ensemble_function} where $d(\cdot,\cdot)$ is the Euclidean dissimilarity, $p=2$ and the weights $w_d$ are all equal.

The soft partition is made crisp by assigning each observation to the cluster with the highest membership probability. This step is not strictly necessary, since one could retain the fuzzy membership structure and work with soft clustering results, as in \cite{maturo2020fuzzy}. We focus on crisp partitions, however, for simplicity and interpretability. This procedure allows to have single partitions based on different number of clusters and to return a final partition whose number of clusters is not necessarily the largest.

The remaining hyperparameters of the proposed methodology that need to be tuned are: the dimension of the projected space $d$, the number of projections $B$ and the size of the ensemble $B^*$. Some indications for these choices, supported by  simulation results, are presented in Section~\ref{parameters}.

\subsection{On the choice of random matrices and ordering criterion}\label{rp_criterion}

The random matrices $\mathbf{A}_b\in\setR^{K\times d}$ that satisfy the JL Lemma can be generated in different ways. We consider two generation processes: \emph{Gaussian} random matrices, whose entries are drawn independently from a $\normal(0,1)$ distribution, with columns subsequently normalized to unit length, and \emph{Haar} matrices, which have orthogonal columns of unit norm and can be generated from the Gaussian random matrices via QR decomposition \citep{haar1933massbegriff}. Although generating Haar matrices is computationally more demanding (involving a QR decomposition as an additional step), its orthogonality ensures that the the geometry of the space is preserved.

The clustering quality criteria employed to order the resulting partitions are measures derived from the resulting mixture of each individual clustering solution. A criterion that ranks the partial clustering results is necessary to discern between partitions that highlight a group structure and to discard uninformative clustering results; an illustrative representation of informative and uninformative projections can be found in Figure~\ref{good_bad}.

In this work, we considered three different measures: the Kullback-Leibler (KL) divergence, the Wasserstein distance and an entropy-based criterion.

The Kullback-Leibler criterion quantifies the dissimilarity between mixture components by means of the KL divergence \citep{kullback1951information}, whose closed-form expression for multivariate normal distributions is the following:
\begin{equation*}
	D_{KL}(\normal_0||\normal_1)=\frac{1}{2}\left[\tr(\boldsymbol{\Sigma}_1^{-1}\boldsymbol{\Sigma_0})-d+(\boldsymbol{\mu}_1-\boldsymbol{\mu}_0)^T\boldsymbol{\Sigma}_1^{-1}(\boldsymbol{\mu}_1-\boldsymbol{\mu}_0) + \log\frac{\det\boldsymbol{\Sigma}_1}{\det\boldsymbol{\Sigma}_0}\right],
\end{equation*}
where $\normal_0$ and $\normal_1$ are multivariate normal distributions with mean vectors $\boldsymbol{\mu}_0$ and $\boldsymbol{\mu}_1$ and covariance matrices $\boldsymbol{\Sigma}_0$ and $\boldsymbol{\Sigma}_1$, respectively. To compute the criterion, we considered the two KL divergences for each pair of mixture components (due to asymmetry) and took their average value. Formally, for the clustering result $b$ with $G$ groups, we can write
\begin{align}\label{kl_criterion}
	KL_b=\frac{\sum_{g=1}^{G}\sum_{h\neq g}D_{KL}(\normal_g||\normal_h)}{G(G-1)}.
\end{align}

Wasserstein distance \citep{kantorovich1942transfer} is used to measure the cost of transporting probability mass between mixture components. Its closed form for the multivariate normal distributions $\normal_0$ and $\normal_1$ is \citep{takatsu2011wasserstein}:
\begin{equation*}
	D_W(\normal_0,\normal_1)=||\boldsymbol{\mu}_0-\boldsymbol{\mu}_1||^2+\tr(\boldsymbol{\Sigma}_0)+\tr(\boldsymbol{\Sigma}_1)-2\tr\left((\boldsymbol{\Sigma}_1^{\frac{1}{2}}\boldsymbol{\Sigma}_0\boldsymbol{\Sigma}_1^{\frac{1}{2}})^{\frac{1}{2}}\right).
\end{equation*}
Unlike the KL divergence, this measure is a proper distance and, thanks to its symmetric structure, its mean can be computed with a shorter set of summands. Thus, the criterion for the clustering result $b$ with $G$ groups can be computed as
\begin{equation*}
	Wass_b=\frac{2\sum_{g=1}^{G}\sum_{h<g}D_W(\normal_g,\normal_h)}{G(G-1)}.
\end{equation*}

The entropy-based criterion evaluates the internal uncertainty of the mixture model, relying on the posterior probabilities associated with the cluster membership of each observation. From its definition in information theory \citep{cover1999elements} , the entropy of the discrete random variable $X_i$, taking values in $\{1,\ldots, G\}$ and representing the cluster membership of unit $i$, can be written as
\begin{equation*}
	H(X_i)=H_i=-\sum_{g=1}^G p_g(x_i)\log(p_g(x_i)),
\end{equation*}
where $p_g(x_i)$ denotes the posterior probability that unit $i$ belongs to cluster $g$, for $g=1,\ldots,G$.
We derive our criterion by computing the mean entropy across all observations and standardising by the maximum possible entropy, i.e. by $\log(G)$. This normalisation is essential to prevent the criterion from favouring solutions with a larger number of clusters. Based on these considerations, we define the entropy-based criterion for clustering result $b$ with $G$ groups as:
\begin{equation*}
	Ent_b=\frac{\sum_{i=1}^{N}H_i}{N\log(G)}.
\end{equation*}
Lower entropy indicates more concentrated components, leading to rankings that favour clusterings with more defined mixture structures.

Each of these criteria captures a different perspective on how to sort clustering results. We propose these alternatives motivated by the intuition that a projection is good if the resulting clusters are well-separated. Additionally, computational cost must be considered, as the criterion has to be evaluated for every projection and for each pair of clusters. Both the KL divergence and the Wasserstein distance meet this requirement in the context of GMMs, since closed-form expressions are available for these measures when the components are multivariate normal distributions.

For the first two criteria, larger values are preferable, whereas for the entropy-based criterion smaller values indicate better clustering performance. Together, these measures provide alternative ways to rank the clustering results, all grounded in the idea that mixture models with well-separated components are more likely to reveal meaningful grouping structures.

\begin{figure}[htbp]
	\centering
	\includegraphics[width=0.95\textwidth]{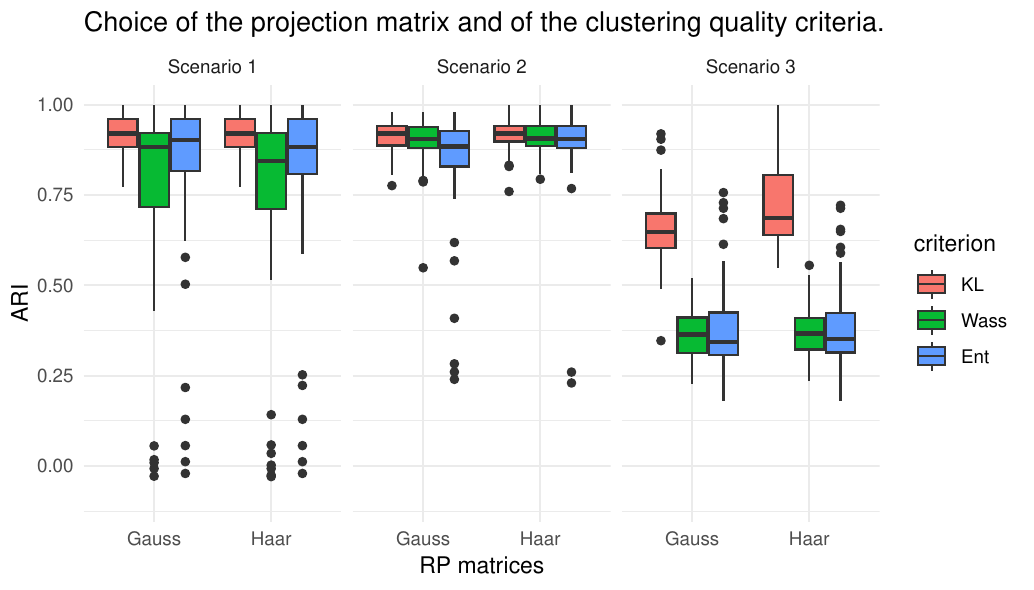}
	\caption{ARI over 100 replicates of the three scenarios for the different random matrices and criteria. Here, $B=1000$, $B^*=100$ and $d$ is equal to $5$, $7$, and $8$ for Scenarios 1, 2 and 3, respectively.}
	\label{fig:boxplot_criterion_rp}
\end{figure}

We assess the impact of the random matrix generation methods and the ordering criteria on clustering performance using the simulation settings described in detail in Section~\ref{simulations}. Performance is evaluated through the adjusted Rand Index \citep[ARI;][]{hubert1985comparing}, computed between the estimated partition and the true cluster labels, over 100 simulated datasets for each scenario. As shown in Figure~\ref{fig:boxplot_criterion_rp}, there is no substantial difference in performance between Gaussian and Haar random matrices. However, results suggest that the KL-based ordering criterion generally yields the best ARI values.

\subsection{On the parameter tuning}\label{parameters}

\begin{figure}[htbp]
	\centering
	\includegraphics[width=0.95\textwidth]{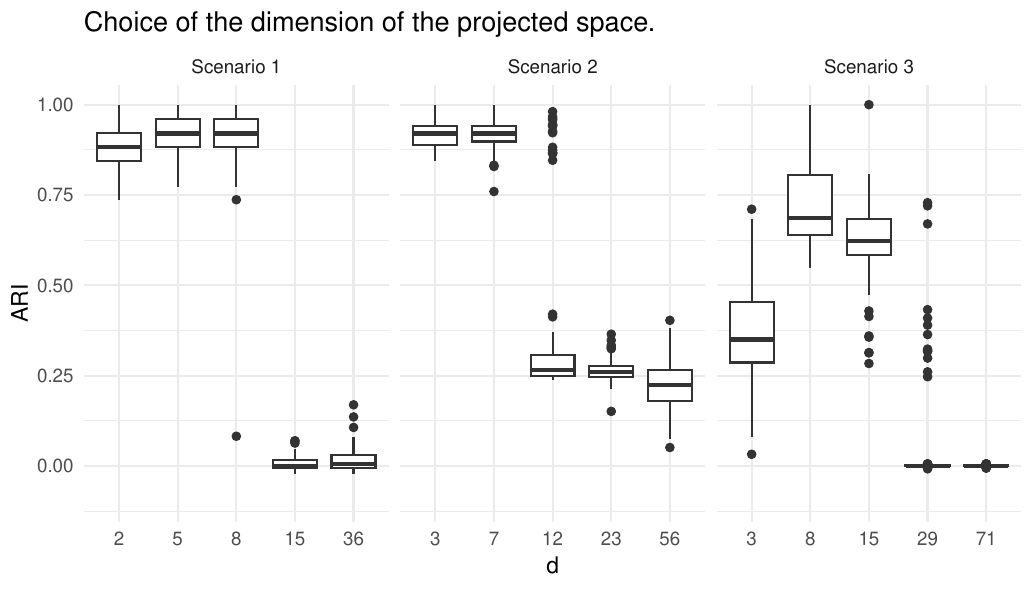}
	\caption{ARI for the three scenarios with different values for $d$. Here $B=1000$, $B^*=100$, the ranking criterion is KL and Haar random matrices are employed.}
	\label{fig:boxplot_d}
\end{figure}

The choice of the parameter $d$ is arguably the most crucial, as the projected dimension must be sufficiently large to preserve the relevant information while remaining small enough to avoid the curse of dimensionality. We explored several values of $d$ defined as

\begin{equation}\label{choice_d}
	d = \lceil a\log(G) \rceil +1,
\end{equation}
with $a \in \{1, 5, 10, 20, 50\}$, following the strategy proposed in \cite{anderlucci2022high} and motivated by the results of \cite{dasgupta1999,dasgupta2000}. Figure~\ref{fig:boxplot_d} reports, for fixed values of $B$ and $B^*$ (set to 1000 and 100, respectively), the boxplots of the ARI obtained for the different choices of $d$ across the three scenarios. In all cases, we observe that too small a dimension fails to recover the true cluster structure, whereas an excessively large dimension leads to a deterioration in performance. The simulations suggest that the optimal choice is $d = \lceil 5\log(G) \rceil + 1$. However, these findings are based on synthetic data and, in the functional setting, the simulated datasets represent a simplified version of real-world structures. For this reason, in practical applications, a choice such as $d = \lceil 10\log(G) \rceil + 1$ may also be appropriate.

\begin{figure}[htbp]
	\centering
	\includegraphics[width=0.95\textwidth]{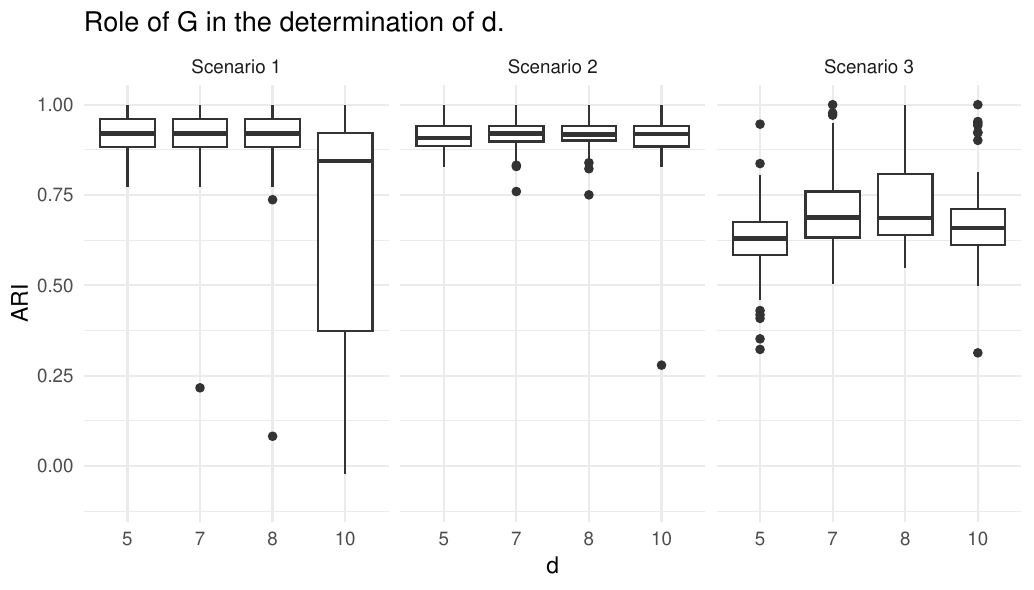}
	\caption{ARI for the three scenarios with different values for $d$ obtained with different values for $G$ in Equation~\ref{choice_d} with $a=5$. Here $B=1000$, $B^*=100$, the ranking criterion is KL and Haar random matrices are employed.}
	\label{fig:choice_d_all}
\end{figure}

The value of $G$ in Equation~\ref{choice_d} for setting $d$ is not used as a decisive parameter; rather, it serves as a heuristic device to guide the choice of the projection dimension. To demonstrate that the tuning of $d$ is not critical and that $G$ does not need to be fixed with high precision, we compare the results of the three scenarios under different values of $d$, computed as $\lceil 5\log(G) \rceil +1$ for various values of $G$. Figure~\ref{fig:choice_d_all} shows substantial robustness to misspecifications of the number of groups. In Scenario~\ref{jacques}, performance remains stable up to $G=5$, corresponding to a projection dimension $d=10=\lceil 5\log(5) \rceil +1$, after which a decline becomes apparent. Results for Scenarios~\ref{ferraty} and \ref{gattone} are stable throughout, with small differences in the latter when $G$ is either too small or too large.

Alternatively, if one wishes to choose the number of dimensions $d$ in a fully data-driven manner, we propose using the entropy on the fuzzy ensemble outputs to identify the most suitable value of $d$ over a predefined set of candidates.

The remaining two parameters, namely $B$ and $B^*$ can be discussed jointly. The performance of the method strongly depends on selecting projections that reveal a clear group structure in the reduced space. Prior studies \citep[see, e.g.][]{kittler2002combining} have shown that ensemble methods are most effective when their constituent members are diverse rather than redundant. RPs naturally induce such diversity by generating perturbed representations of the data; however, including an excessive number of projections may be detrimental and increases computational burden. Conversely, using too few projections may not provide the ensemble with sufficient information to yield reliable results.

\begin{figure}[htbp]
	\centering
	\includegraphics[width=0.95\textwidth]{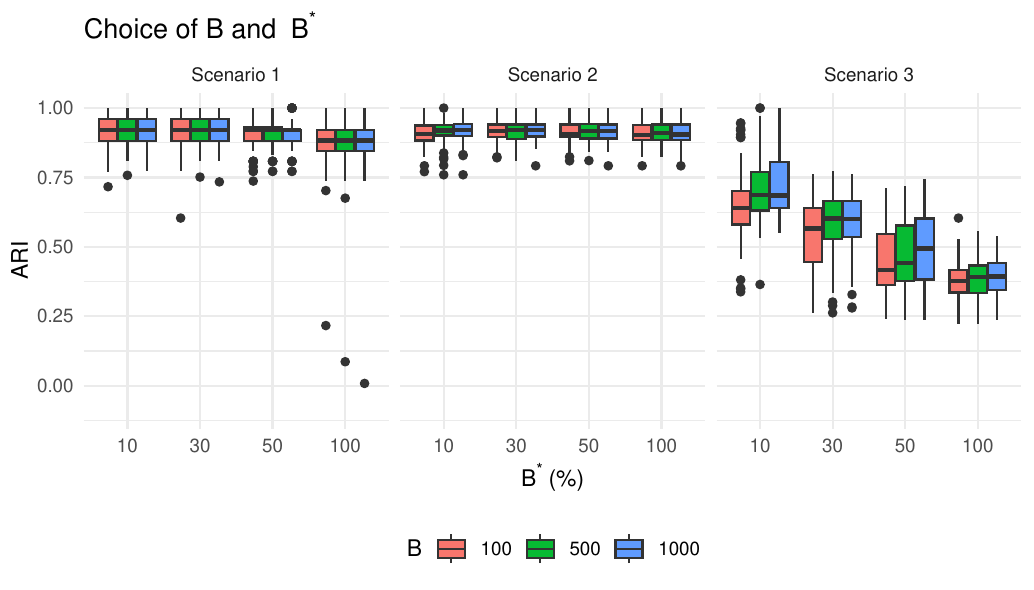}
	\caption{ARI results for the three scenarios under different combinations of $B$ and $B^*$. The value of $B^*$ is expressed as a percentage of $B$. Here, $d$ is set to $5$, $7$ and $8$ for the three scenarios, respectively; the ranking criterion is KL and Haar random matrices are employed.}
	\label{fig:boxplot_B_Bstar}
\end{figure}

We examined three different values of $B$ ($100$, $500$, and $1000$) while varying $B^*$ as a percentage of $B$, specifically $10\%$, $30\%$, $50\%$, and $100\%$. In this analysis, the projection dimension $d$ was fixed at $5$, $7$, and $8$ for the three scenarios, corresponding to $G=$ 2, 3 and 4, respectively. As expected, Figure~\ref{fig:boxplot_B_Bstar} shows that, all else being equal, increasing the number of random projections improves clustering performance. With respect to $B^*$, the results indicate that retaining as little as 10\% of the projections already achieves satisfactory performance; increasing this proportion tends to introduce non-informative projections, which may ultimately degrade the overall results.

\section{Simulation study}\label{simulations}
The performance of the proposed algorithm is evaluated in a variety of scenarios through an extensive simulation study, where the number of clusters and the degree of their overlapping vary. Observations for each cluster are generated according to the model described in Equation~\ref{fundata_model}. In particular, the underlying function $x$ is built from one or a combination of two analytic functions. Three artificial setups are discussed:
\begin{enumerate}
	\item\label{jacques} $G=2$ groups, $n=1001$ equidistant sampling points $t \in [1,21]$ and $N=100$ observed curves, inspired by the data generating structure presented in \cite{jacques2013funclust}. The underlying functions are:
	\begin{align*}
		x_1(t)&=U_1h_1(t)+U_2h_2(t), \\
		x_2(t)&=U_1h_1(t), \\
	\end{align*}
	where $U_1$ and $U_2$ are independent Gaussian variables such that $\E[U1]=E[U_2]=0$ and $\Var(U_1)=\Var(U_2)=1/12$, while
	\begin{align*}
		h_1(t)&=\max(6-|t-7|,0),\\
		h_2(t)&=\max(6-|t-15|,0).\\
	\end{align*}
	\item\label{ferraty} $G=3$ groups, $n=1001$ equidistant sampling points $t \in [1,21]$ and $N=150$. This is similar to the data generated in \cite{ferraty2003curves}. Here, the underlying functions for the three groups are:
	\begin{align*}
		x_1(t)&=Uh_1(t)+(1-U)h_2(t),  \\
		x_2(t)&=Uh_1(t)+(1-U)h_3(t),  \\
		x_3(t)&=Uh_2(t)+(1-U)h_3(t), \\
	\end{align*}
	where $U$ is the uniform random variable on $(0,1)$ and the $h_g$ are the shifted triangular waveforms:
	\begin{align*}
		h_1(t)&=\max(6-|t-11|,0),\\
		h_2(t)&=h_1(t-4),\\
		h_3(t)&=h_1(t+4).\\
	\end{align*}
	\item\label{gattone} The third example is generated in the same way of Scenario $3$ in \cite{gattone}, with $G=4$ groups, $n=101$ equidistant sampling points $t \in [-1,1]$ and $N=100$. Specifically, we have that:
	\begin{align*}
		x_1(t)&=h_1(t), \\
		x_2(t)&=h_2(t), \\
		x_3(t)&=h_1(t)+h_2(t),\\
		x_4(t)&=h_2(t)-h_1(t), \\
	\end{align*}
	with auxiliary functions
	\begin{align*}
		h_1(t)&=\cos(20t),\\
		h_2(t)&=\sin(20t).\\
	\end{align*}
\end{enumerate}

The first two scenarios are designed to simulate high frequency functional data, with a number of observed points on the order of $10^3$. Scenario~\ref{gattone}, by contrast, represents a setting with a substantially smaller number of sampling points.

In all scenarios, each unit has equal probability of belonging to any cluster. The curves are observed at equidistant points: in Scenarios \ref{jacques} and \ref{ferraty} the elements of vector $\mathbf{t}$ lie in the interval $[1,21]$, while in Scenario \ref{gattone} the interval is $[-1,1]$.

The error vector $\mathbf{e}$ is generated from a multivariate Gaussian distributions with mean vector $\mathbf{0}\in \setR^n$ and covariance matrices $\sigma^2\mathbf{I}_n$, where $\mathbf{I}_n\in\setR^{n\times n}$ is the identity matrix. The noise variances are set to $\sigma^2=$ 1/12, 1 and 1/25 in Scenarios \ref{jacques}, \ref{ferraty} and \ref{gattone}, respectively. For each setting, 100 replicates were generated.

\begin{figure}
	\centering
	\includegraphics[width=\textwidth]{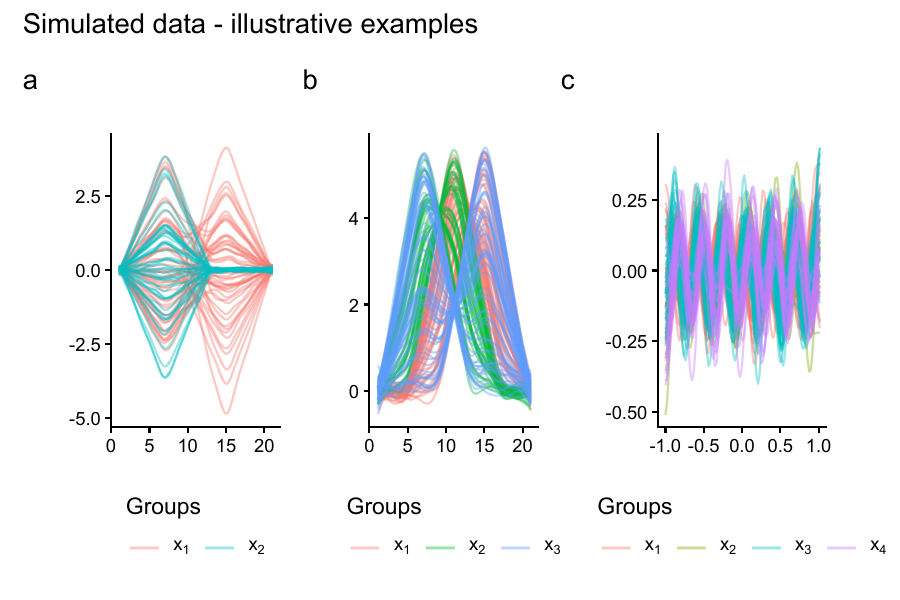}
	\caption{Examples of simulated data according to the three different scenarios: panel (a) depicts scenario \ref{jacques}, panel (b) displays scenario \ref{ferraty} and panel (c) illustrates scenario \ref{gattone}. Different colours reflect different cluster membership. The curves are smoothed using cubic splines penalized via an integrated squared second derivative term.}
	\label{fig:scenarios}
\end{figure}

The curves were smoothed using B-spline basis of order $4$ (cubic splines) penalized via an integrated squared second derivative term. The same number of basis functions $K$ and the same penalty parameter $\lambda$ were used across the $100$ replications of each scenario. The GCV procedure suggests $K=100$ and $\lambda=1$ for Scenario \ref{jacques}, and $K=200$ with $\lambda=10$ for Scenario \ref{ferraty}. In Scenario \ref{gattone}, where the number of sampling points is only $101$ and thus fewer basis functions would typically suffice, we deliberately set $K$ equal to the number of sampling points to test the robustness of our model to potential overfitting, while $\lambda$ was selected according to the GCV recommendation and set to $10^{-4}$. A summary of the dataset characteristics and GCV parameters setting is provided in Table~\ref{summary_table}, while Figure~\ref{fig:scenarios} illustrate a smoothed example dataset for each scenario.

\begin{table}[htbp]
	\centering
	\caption{Summary of the parameters used in each simulation for data generation and basis function representation. $t_0$ and $t_n$ are the first and last elements of the vector of time points $\mathbf{t}$. The value for $\sigma^2$ refers to the variance of the error term.}
	\begin{tabular}{cccccccc}
		\toprule
		Scenario & $G$ & $n$ & $N$ & $[t_0,t_n]$ & $\sigma^2$ & $K$ & $\lambda\phantom{11}$ \\
		\midrule
		\ref{jacques} & $2$ & $1001$ & $100$ & $[1,21]$ & $\frac{1}{12}$ & $100$ & $\phantom{0}1\phantom{^{-4}}$ \\
		\ref{ferraty} & $3$ & $1001$ & $150$ & $[1,21]$ & $1$ & $200$ & $10\phantom{^{-4}}$ \\
		\ref{gattone} & $4$ & $\phantom{1}101$ & $101$ & $[-1,1]$ & $\frac{1}{25}$ & $101$ & $10^{-4}$ \\
		\bottomrule
	\end{tabular}
	\label{summary_table}
\end{table}

All the analysis were conducted using \textsf{R} \citep{R-base}. In particular, for each reduced dataset $\mathbf{X}_b$, the GMM are fitted using the \texttt{mclust} package \citep{mclust}, while the consensus clustering is obtained through the \texttt{cl\_consensus} function in the \texttt{clue} package \citep{clue}. The number of clusters in \texttt{Mclust} is allowed to vary from $2$ to $9$. In the simulation study, to alleviate computational difficulties (both in terms of runtime and occasional numerical issues arising in specific cases), the covariance structure is selected from all available models except \texttt{VVE} and \texttt{EVE}, corresponding respectively to ellipsoidal models with equal volume and ellipsoidal models with equal volume and orientation. For \texttt{cl\_consensus}, the argument specifying the consensus algorithm is ``\texttt{SE}'', which denotes the Soft least squared Euclidean consensus.

For each scenario we ran the algorithm using Haar random matrices and KL criterion with parameters $B=1000$, $B^*=100$ and $d=\lceil 5\log(G) \rceil +1$. We compared our model with alternatives from the literature (see Table~\ref{sota}) using the adjusted Rand index.

\begin{table}[htbp]
	\centering
	\caption{Literature methods and corresponding references}
	\begin{tabular}{lcl}
		\toprule
		Method name or acronym & Method category& Reference \\
		\midrule
		RFKMn & Adaptive & \cite{gattone} \\
		funHDDC & Adaptive & \cite{bouveyron2011model} \\
		Funclust & Adaptive & \cite{jacques2013funclust} \\
		SPFC-d$_{\omega}$ & Adaptive & \cite{yu2025distance} \\
		SaS-Funclust & Adaptive & \cite{centofanti2024sparse} \\
		EMCluster & Filtering & \cite{Chen2015EMClusterpackage} \\
		FADPclust  & Filtering & \cite{ren2023multivariate} \\
		gmfd\_km & Filtering & \cite{martino2019k} \\
		\bottomrule
	\end{tabular}
	\label{sota}
\end{table}

For convenience, we refer to our proposed method as FunCluRPE. For all competing methods from the literature, the number of clusters $G$ was assumed to be known, and the hyperparameters were set according to the recommendations provided in the original publications. Since RFKMn and SPFC-d$_\omega$ do not include built-in routines for selecting the parameters $Q$ and $\omega$, respectively, we report the best ARI results obtained after testing multiple parameter configurations.

\begin{figure}[htbp]
	\centering
	\includegraphics[width=0.95\textwidth]{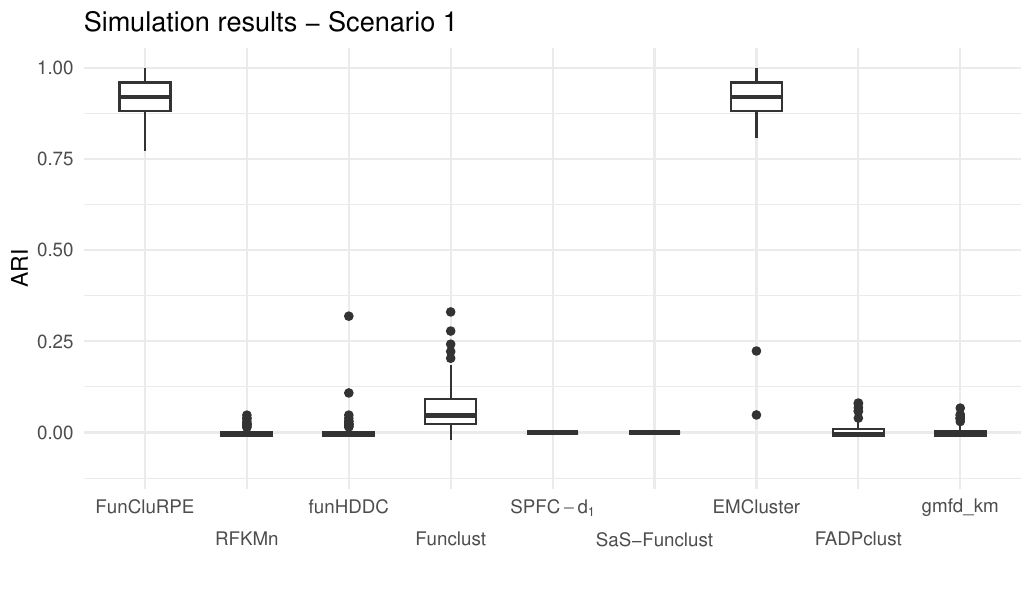}
	\caption{Comparison of ARI performances for Scenario \ref{jacques}.}
	\label{scenario1_comparison}
\end{figure}

\begin{figure}[htbp]
	\centering
	\includegraphics[width=0.95\textwidth]{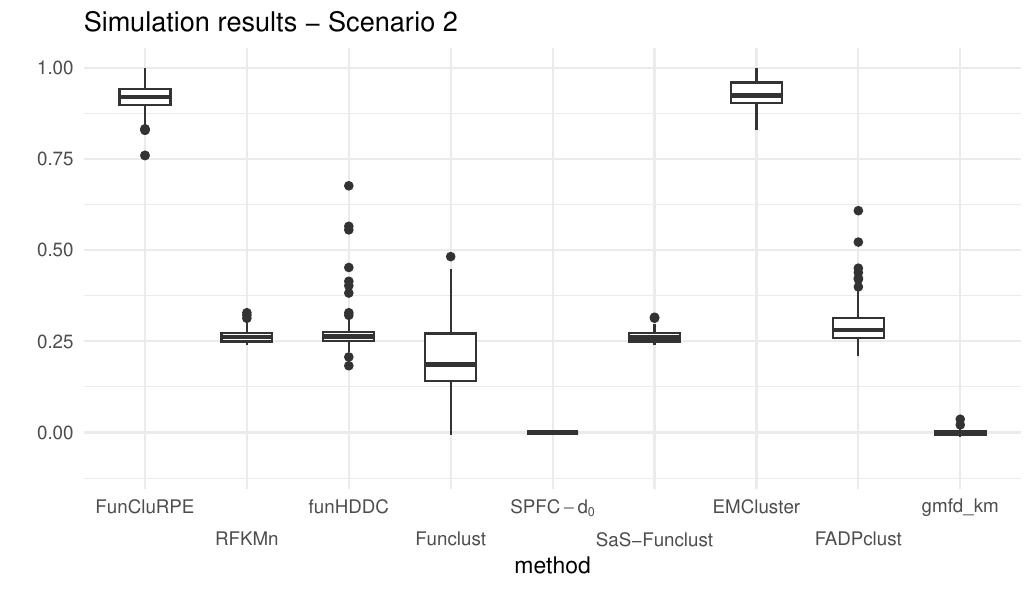}
	\caption{Comparison of ARI performances for Scenario \ref{ferraty}.}
	\label{scenario2_comparison}
\end{figure}

In Scenarios \ref{jacques} and \ref{ferraty}, the proposal FunCluRPE performs well, achieving results comparable to those of EMCluster, which is the best-performing method among the literature models considered (see Figures \ref{scenario1_comparison} and \ref{scenario2_comparison}).

\begin{figure}[htbp]
	\centering
	\includegraphics[width=0.95\textwidth]{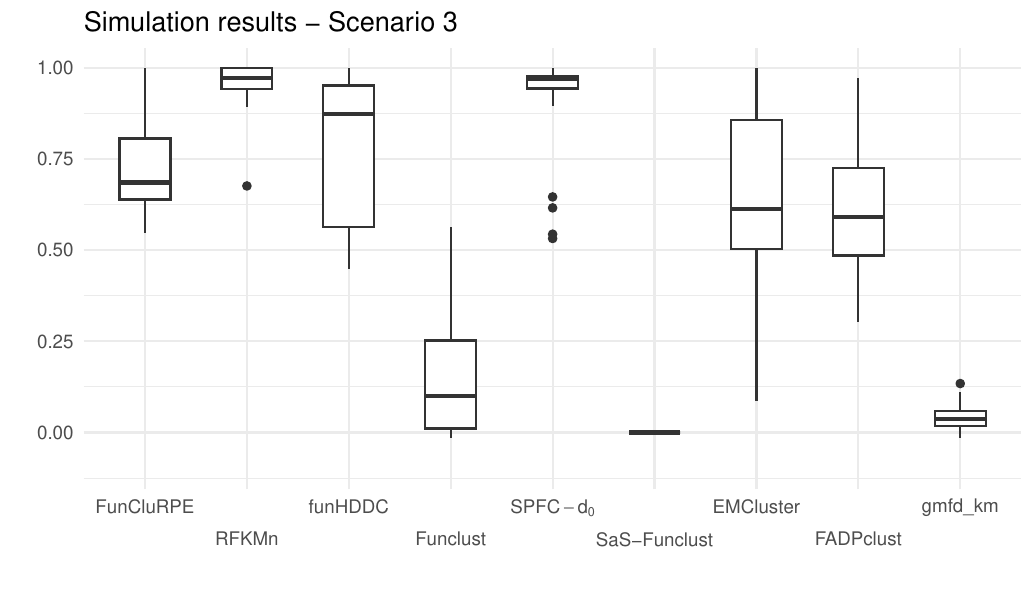}
	\caption{Comparison of ARI performances for Scenario \ref{gattone}.}
	\label{scenario3_comparison}
\end{figure}

In Figure~\ref{scenario3_comparison}, results for Scenario \ref{gattone} are reported, where FunCluRPE performs comparably to funHDDC, and slightly worse than RFKMn and SPFC-d$_{\omega}$. It is worth noticing, however, that this simulation is inspired by the data generating process introduced in \cite{gattone}, where RFKMn was originally proposed. Moreover, the reported results for SPFC-d$_{\omega}$ correspond to $\omega=0$ thereby discarding all information contained in the derivative curves, which constituted the main innovation of that model.

Another noteworthy aspect of the proposed method is that it does not require specifying the number of clusters in advance. Instead, the ensemble procedure automatically determines the most suitable number of groups for each dataset. To evaluate performance in this respect, Table~\ref{clusters_table} reports the number of clusters identified across simulations. As shown, the method correctly recovers the true number of clusters in most cases.

\begin{table}[h]
	\caption{Number of clusters identified across the $100$ simulations for each scenario.}
	\label{clusters_table}
		\begin{tabular*}{\textwidth}{@{\extracolsep\fill}lcccccc}
		\toprule%
		& \multicolumn{2}{@{}c@{}}{Scenario \ref{jacques} (true $G=2$)} & \multicolumn{2}{@{}c@{}}{Scenario \ref{ferraty} (true $G=3$)}& 		\multicolumn{2}{@{}c@{}}{Scenario \ref{gattone} (true $G=4$)} \\\midrule%
		$G$ & $2$  & $3$ & $3$ & $4$ & $3$ & $4$ \\
		\cmidrule{2-3}\cmidrule{4-5}\cmidrule{6-7}
		& $\mathbf{99}$ & $1$  & $\mathbf{99}$ & $1$ & $46$ & $\mathbf{54}$\\
		\bottomrule
	\end{tabular*}
\end{table}

\section{Real data applications}\label{real_data}
We evaluate the proposed method on two real datasets. As the computational burden is substantially lower than in the simulation study, we allow the \texttt{Mclust} function to select among all covariance structures, including those previously excluded (\texttt{VVE} and \texttt{EVE}). We also consider a variant of our approach in which the true number of clusters is supplied to \texttt{Mclust} (denoted FunCluRPE\_G). All experiments were executed on a server, equipped with two AMD EPYC 7763 processors (128 total cores), 2 TB of RAM, and running SUSE Linux; computations for the proposal were parallelised and performed using 8 cores in parallel.

\begin{figure}[htbp]
	\centering
	\includegraphics[width=0.95\textwidth]{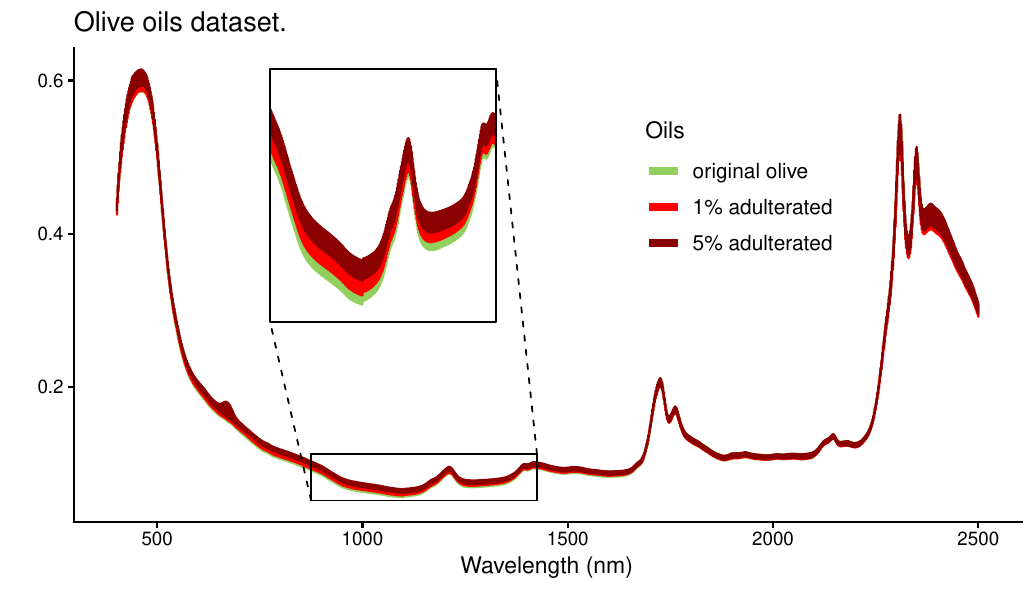}
	\caption{Plot of the olive oils dataset. The curves are coloured according to the level of adulteration. For clarity, a zoomed-in view of the spectral range between $900$ nm and $1400$ nm is shown.}
	\label{oil_figure}
\end{figure}

The first dataset consists of $46$ Greek authentic extra-virgin olive oil samples collected by \cite{downey2002detecting}. Visible and near-infrared spectra were recorded over the wavelength range of $400-2498$ nm ($400-750$ nm for the visible region and $1100-2498$ nm for the near-infrared region) at a sampling rate of $2$ nm, resulting in $n=1050$ measurement points. Each sample was analysed in its pure form and after adulteration with $1\%$ and $5\%$ authentic sunflower oil, resulting in a total of $N=138$ spectral samples and $G=3$ groups. The aim is to uncover the latent grouping structure, with particular interest in distinguishing the unadulterated samples from the contaminated ones. Using B-spline basis function of order $4$, the GCV procedure selects $K=1050$ and $\lambda=10^{-5}$. A representation of the data under this setting is displayed in Figure~\ref{oil_figure}. We apply FunCluRPE with $d=\lceil 10\log(3) \rceil +1=12$, $B=1000$ $B^*=100$, Gaussian random matrices and KL criterion. The resulting ARI values and computational times are summarised in Table~\ref{oils_table}.
Providing the correct number of clusters (FunCluRPE\_G) does not improve accuracy; also, even without this information the method successfully identifies the true number of groups.
The substantial reduction in dimensionality (from $1050$ to $12$), combined with the use of Gaussian projections and the relatively small sample size, contribute to the high computational efficiency observed in this example. Moreover, the confusion matrix in Table~\ref{oils_confusion} shows that FunCluRPE is able to distinguish clearly between the authentic oils from the adulterated ones.

\begin{table}[htbp]
	\centering
	\caption{ARI and computation time (in seconds) of FunCluRPE, FunCluRPE\_G and the methods in Table~\ref{sota} for the olive oils data.}
	\begin{tabular}{lcr}
		\toprule
		Method & ARI & time (s)\\
		\midrule
		FunCluRPE & $0.82$ & $77$ \\
		FunCluRPE\_G & $0.73$ & $16$ \\
		RFKMn & $0.25$ & $4296$ \\
		funHDDC & $0.27$ & $18020$ \\
		Funclust &  $0.00$ & $1413$ \\
		SPFC-d$_{0.50}$ & $0.00$ & $20214$ \\
		SaS-Funclust & $0.00$ & $1239$ \\
		EMCluster & $0.29$ & $4975$ \\
		FADPclust & $0.26$ & $51$ \\
		gmfd\_km & $0.26$ & $296$ \\
		\bottomrule
	\end{tabular}
	\label{oils_table}
\end{table}

\begin{table}
	\centering
	\caption{Classification table for FunCluRPE for the olive oil data.}	
	\begin{tabular}{lccc}
		\toprule
		\multirow{2}*{Cluster} & \multicolumn{3}{c}{Oil label} \\
		\cmidrule(lr){2-4}
		& 0\% & 1\% & 5\% \\
		\midrule
		1 & $0$ & $0$ & $37$ \\
		2 & $0$ & $46$ & $9$ \\
		3 & $46$ & $0$ & $0$ \\
		\bottomrule
	\end{tabular}
	\label{oils_confusion}
\end{table}

\begin{figure}[htbp]
	\centering
	\includegraphics[width=0.95\textwidth]{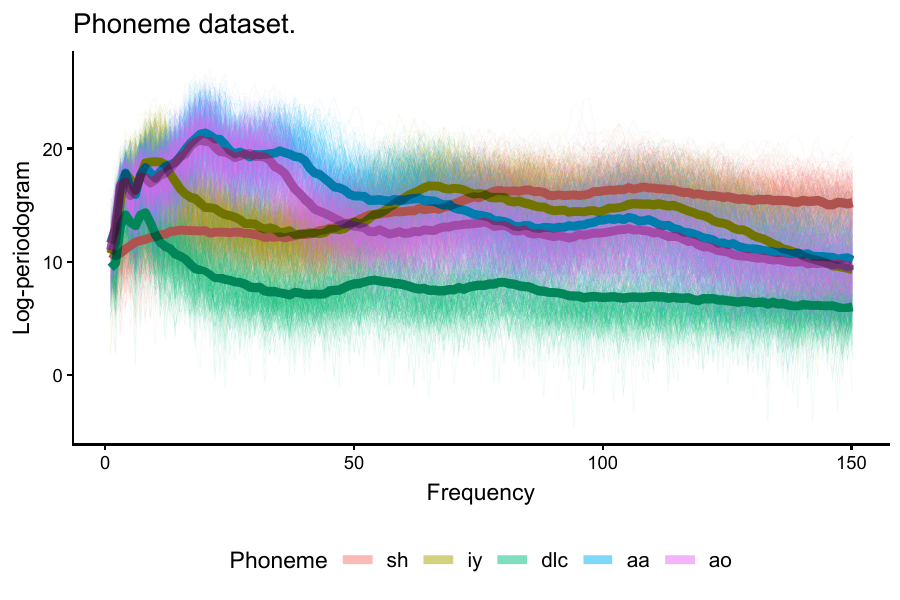}
	\caption{Plot of the phoneme dataset. Mean curves for each phoneme are highlighted in bold.}
	\label{phoneme_figure}
\end{figure}

A second example, taken from the supplementary material of \cite{gattone}, concerns a phoneme classification problem. The data consist of log-periodograms of speech recordings, each of $32$ ms duration, corresponding to $G=5$ phoneme classes: ``sh'' (as in she), ``dcl'' (as in dark), ``iy'' (the vowel in she), ``aa'' (the vowel in dark), and ``ao'' (the first vowel in water). Following \cite{ferraty2003curves}, we consider only the first $n=150$ frequencies, obtained from signals recorded at a $16$ kHz sampling rate. Each phoneme class includes $400$ recordings, resulting in a dataset of $N=2000$ log-periodograms with known class membership (see Figure~\ref{phoneme_figure}). The GCV procedure, applied to B-spline basis of order 4, selects $K=150$ and $\lambda=10^{-2}$. FunCluRPE model was run with Gaussian matrices, KL criterion and parameters $d=\lceil 10\log(5) \rceil +1=18$, $B=1000$ and $B^*=100$. As shown in Table~\ref{phoneme_table}, the proposed method performs well relative to competing approaches from the literature. The confusion matrix in Table~\ref{phoneme_confusion} further demonstrates the effectiveness of FunCluRPE in correctly identifying most classes. Distinguishing between the phonemes ``aa" and ``ao" is particularly challenging; indeed, Figure~\ref{phoneme_figure}, indicates that they exhibit the greatest degree of overlap.

\begin{table}[htbp]
	\centering
	\caption{ARI and computation time (in seconds) of FunCluRPE, FunCluRPE\_G and the methods in Table~\ref{sota} for the phoneme data.}
	\begin{tabular}{lcr}
		\toprule
		Method & ARI & time (s)\\
		\midrule
		FunCluRPE & $0.76$ & $1978$ \\
		FunCluRPE\_G & $0.76$ & $637$ \\
		RFKMn & $0.74$ & $274$ \\
		funHDDC & $0.60$ & $16716$ \\
		Funclust &  $0.11$ & $1312$ \\
		SPFC-d$_{0.75}$ & $0.63$ & $1401$ \\
		SaS-Funclust & $0.72$ & $9963$ \\
		EMCluster & $0.69$ & $324$ \\
		FADPclust & $0.77$ & $70$ \\
		gmfd\_km & $0.43$ & $898$ \\
		\bottomrule
	\end{tabular}
	\label{phoneme_table}
\end{table}

\begin{table}
	\centering
	\caption{Classification table for FunCluRPE for the phoneme data.}	
	\begin{tabular}{lccccc}
		\toprule
		\multirow{2}*{Cluster} & \multicolumn{5}{c}{Phoneme label} \\
		\cmidrule(lr){2-6}
		& sh & iy & dlc & aa & ao \\
		\midrule
		1 & $0$ & $0$ & $0$ & $381$ & $393$ \\
		2 & $0$ & $399$ & $12$ & $0$ & $1$ \\
		3 & $0$ & $1$ & $387$ & $0$ & $0$ \\
		4 & $0$ & $0$ & $0$ & $19$ & $6$ \\
		5 & $400$ & $0$ & $1$ & $0$ & $0$ \\
		\bottomrule
	\end{tabular}
	\label{phoneme_confusion}
\end{table}

The nice computational advantage of FunCluRPE, that was exploited both in simulations and in the examples, is that if can be easily parallelised, reducing its computational time.

\section{Discussion}\label{conclusions}

The problem of clustering functional data, where each observation is represented as a curve varying continuously over time, is not trivial. The proposed methodology exploits the Random Projections framework, which simultaneously perform dimensionality reduction and generates multiple stochastic views of the data. Each projection is independently clustered, and the resulting partitions are subsequently combined through an ensemble procedure to obtain a robust consensus solution.

Extensive testing of different algorithmic configurations showed that the best-performing setting employs Gaussian Mixture Models as the base clustering algorithm and the Kullback-Leibler divergence as ordering criterion for the random projections. This combinations consistently yields accurate and stable clustering results across both simulated and real data.

A key strength of the proposed framework is its data-driven adaptability. Unlike many traditional clustering methods, it does not require the number of clusters to be specified a priori: the structure of the data and the ensemble mechanism jointly determine an appropriate partitioning. Moreover, the method allows for the direct use of GCV results to obtain the basis coefficients, alleviating concerns about the dimensionality of coefficient spaces and simplifying implementation and interpretation. Another noteworthy aspect is the robustness of the tuning parameters: the performance of the model exhibits little sensitivity to variations in $B$, $B^*$ and $d$, suggesting that the procedure remains stable across a broad range of settings. The R code of FunCluRPE is available at \url{https://github.com/mmori8/FunCluRPE}.

Future work will focus on extending the methodology to multivariate functional data, enabling the simultaneous analysis of several functional variables, while accounting for their dependence structure. This extension presents additional methodological and computational challenges and warrants further investigation.

\section*{Acknowledgments}
Matteo Mori received a PhD scholarship financially supported by the European Union---NextGenerationEU, under the Italian National Recovery and Resilience Plan (PNRR), Mission 4, Component 2, Investment 3.3, as part of the research fellowship Ex D.M. 352/2022 and by Arithmos Srl. \\
Laura Anderlucci declares no financial interests.

\noindent We gratefully acknowledge Prof. Gerard Downey for sharing the oil data used in this study.

\bibliographystyle{chicago}
\bibliography{manuscript_biblio}

\end{document}